# Toward Development of Machine Learned Techniques for Production of Compact Kinetic Models

Mark Kelly[1], Mark Fortune[1], Gilles Bourque[2,3], Stephen Dooley[1]

[1]School of Physics, Trinity College Dublin, Dublin, Ireland

[2]Siemens Energy Canada Ltd, Montréal, QC H9P 1A5, Canada

[3]McGill University, Montréal, QC H9P 1A5, Canada

**Abstract**

Chemical kinetic models are an essential component in the development and optimisation of combustion devices through their coupling to multi-dimensional simulations such as computational fluid dynamics (CFD). Due to the significant level of detail contained within, detailed chemical kinetic models are computationally prohibitive for use in CFD. Therefore, low-dimensional kinetic models which retain good fidelity to the reality are needed, the production of which requires considerable human-time cost and expert knowledge.

Here, we present a novel automated compute intensification methodology to produce overly-reduced and optimised ("compact") chemical kinetic models. This coded algorithm, termed Machine Learned Optimisation of Chemical Kinetics (MLOCK), systematically perturbs each of the four sub-models of a chemical kinetic model to discover what combinations of terms results in an objectified good model. A virtual reaction network comprised of $n$ species is first obtained using conventional mechanism reduction procedures. Once $n$ is lower than a threshold value, the model performance is typically poor. To counteract this, the weights (virtual reaction rate constants) of important connections (virtual reactions) between each node (species) of the virtual reaction network are numerically optimised to replicate selected calculations across four sequential phases.

The first version of MLOCK, (MLOCK1.0) simultaneously perturbs all three virtual Arrhenius reaction rate constant parameters for important connections and assesses the suitability of the new parameters through objective error functions, which quantify the error in each compact model candidate's calculation of the optimisation targets, which may be comprised of detailed model calculations and/or experimental data.

In this study, the MLOCK algorithm is demonstrated by automatically creating compact models for the archetypal case of methane air combustion. It is shown that the NUGMECH1.0 detailed model comprised of 2,789 species is reliably compacted to 15 species (nodes), whilst retaining an overall fidelity of ~87% to the detailed model calculations, outperforming the prior state-of-art.

# 1. Introduction

Due to the direct link between greenhouse gas emissions and global warming, increasingly stringent legislative demands are being placed on emission efficiency performances of combustion devices [8]. As a result, engine manufacturers must advance their technology through increasing power-emission efficiencies whilst maintaining an acceptable level of operability.

To find the burner geometry and operating conditions that maximises the emission efficiency, a greater understanding of the combustion phenomena occurring inside the combustor is needed. Computer reacting flow simulations such as computational fluid dynamics (CFD) and chemical reacting networks (CRN) are used as rapid development tools in the production of these new engines as they provide quick and efficient results compared to real life experiments and can be performed at operating conditions that are both challenging and expensive to perform physically. These simulations are performed by applying mathematical solvers to solve governing equations relevant to a specific reactor geometry. These simulations describe the evolution of the gas fuel mixture, providing information on the efficiency of the combustion process at the specific set of operating conditions.

To ensure the results from these simulations are reliable, an accurate description of the pertinent chemistry must be provided. This information is supplied in the form of a chemical kinetic model.

A chemical kinetic model provides a mathematical description of a chemical reaction mechanism. These models typically comprise four components:

1. Reaction mechanism (or reaction network) which lists the elements, chemical species, and reactions to be considered.
2. Kinetics component containing Arrhenius reaction rate constant parameters ($A$, $n$, and $E_A$) for each reaction which are used to calculate the rate at which each reaction proceeds as a function of temperature, pressure, and reactant concentration.
3. Molecular thermodynamics descriptors for each species with which the changing thermodynamic state of the reacting gas and reaction equilibria can be calculated.
4. Molecular transport descriptors for each species which describe the transport of mass and energy across the changing state of the gas.

Over the years, the Chemkin [12] format of these descriptors has grown to become the community standard format compatible with the computational methods of both developers and users in commercial CFD codes.

In these types of software, the information contained in the four components above are used as inputs in the mathematical solver's governing equations, which include conservation of elemental mass, energy, speciation, and momentum. These are solved at each time and/or spatial point to calculate the state of the fuel mixture as it evolves in the reactor. The rate at which each reaction in the reaction network proceeds is calculated by computing the concentration of each species in the system and the reaction rate, *K* of each reaction:

$$K_i = k_{fi} \prod_{j=1}^{Ns} [X_j]^{\upsilon'_{ji}} - k_{ri} \prod_{j=1}^{Ns} [X_j]^{\upsilon''_{ji}} \quad (1)$$

Where $[X_j]$ is the molar concentration of species *j*, $\upsilon'_{ji}$ and $\upsilon''_{ji}$ are the forward and reverse stoichiometric coefficients of the *j*'th species in the *i*'th reaction with forward and reverse reaction rate constants, $k_{fi}$ and $k_{ri}$ respectively. The forward reaction rate constants, $k_f$ are computed using the Arrhenius rate constant parameters provided for each reaction, in the kinetic model:

$$k_{fi} = A_i T^{n_i} e^{\left(-\frac{E_{A,i}}{RT}\right)} \quad (2)$$

Where *A* is the pre-exponential factor; *n* is the temperature exponent; $E_A$ is the activation energy, *R* is the universal gas constant, and *T* is the temperature. The reverse reaction rate constants, $k_r$ are computed through solving the equilibrium constant for each reaction, using information provided in the thermodynamics component of the kinetic model. In transient solutions, the diffusion behaviour of each species is additionally considered through the computation of diffusion coefficients, diffusion velocities, thermal conductivities, and thermal diffusion coefficients. These are computed by solving a system of equations involving binary diffusion coefficients, species mole fractions, and thermodynamic and molecular properties of the species. The data contained in the thermodynamics and transport data components of the chemical kinetic model are used as inputs for these equations. These additional transport equations significantly increase the computational cost of the simulation. Lu et al. [15]

demonstrated that the computational cost of performing a transient simulation scales with the number of species in the kinetic model, $N_s$ as:

$$C \propto \alpha N_s + \beta N_s^2 + \gamma N_s^3 \qquad (3)$$

Where α, β, and γ are coefficients for the rate evaluation, species diffusion, and Jacobian factorisation respectively. These coefficients are simulation-specific and depend largely on the degree of spatial and time resolution imposed on the solver. This non-linearity between number of species and computational cost presents difficulties when performing simulations.

Detailed chemical kinetic models are conceptually attractive as they are almost completely condensable to chemical reaction theory. Their elaboration over the past few decades has been encouraged by the maturation of computational tools that implement electronic structure (density functional theories) and other chemical theories (transition state theory) to determine much of the reaction mechanism and reaction kinetic components to close to quantitative accuracy. However, a lesson from this dramatic progress has been a questioning of the cost-benefit trade-off of the human time taken to gather this detail, and the computational time needed to solve it, versus the knowledge value offered to combustor design. Due to the complexity involved in reacting flow simulations, kinetic models in excess of one hundred species are of limited utility. Thus, to allow for rapid iterative deployment of multi-dimensional reacting flow simulations in combustor design process, a much smaller kinetic model of reduced dimensionality ($N_s$) is needed, however, this must retain a predictive fidelity in the computation of key quantities of interest, with regard to a fully detailed description. This has been the purpose of mechanism reduction methods.

**1.1 Model Reduction**

Model reduction (or mechanism reduction) is the removal of low importance species and reactions from the reaction network of a detailed kinetic model, whilst not significantly diminishing the fidelity of the resulting model's calculations. The goal is to find the best compromise between computational cost and fidelity of the resulting calculations. Methodologies to achieve this have received significant attention and are a mature field being reviewed several times [15, 19, 20]. However, mechanism reduction methodologies are intrinsically limited, because as species are removed from the reaction network, the fidelity of the calculations decreases until a critical point of necessary detail is reached. After this point, further removal of species leads to a deterioration in the model performance. This is due to the over-reduction of the mechanism, removing species that are important in mediating the reaction flux through the system from reactants to products. The removal of these important pathways results in a severe compromising of the authentic reaction flux to the reality. This threshold is the limit of physically authentic kinetic models produced through conventional mechanism reduction techniques. Figure 1 shows this limit to be approximately 25 species for the example of the methane system across the specific performance envelope. This behaviour was tested using kinetic models resulting from two separate reduction methodologies to illustrate that this coupling is not unique to a specific mechanism reduction technique. The more complex the target set, over the wider range of conditions, the larger the number of species required to retain an accurate fundamental description of the reaction mechanism. However, kinetic models of this size are too large for efficient implementation in CFD. Therefore, kinetic models must be produced that exist below this threshold whilst still maintaining high fidelity to the target combustion property set.

One method for achieving this is to obtain a skeletal model from conventional mechanism reduction and to then apply analytical techniques such as quasi-steady state (QSSA) and partial equilibrium assumptions [23, 24] to solve specific species in the reacting flow in terms of algebraic equations, thereby removing them from the transport equations, significantly reducing computational cost. However, use of these analytical techniques limits the model's deployment due to the requirement of additional kinetic sub-routines to be employed in industry-standard packages such as ChemKin.

**1.2 Compact Kinetic Models**

The concept of a compact kinetic model is an attractive alternative concept that has the potential to meet this challenge. In a compact model, all detail that is not absolutely necessary to the calculation of a defined set of performance calculations is removed. The errors that result owing to the removal of this detail are compensated for by optimisation of any of the reaction kinetic, molecular thermodynamic or mass/energy transport terms comprising the model. As compact models do not contain the necessary level of detail in the reaction network component to fundamentally describe the reaction mechanism, the description of the reacting flow is virtual rather than attempting to be physically authentic.

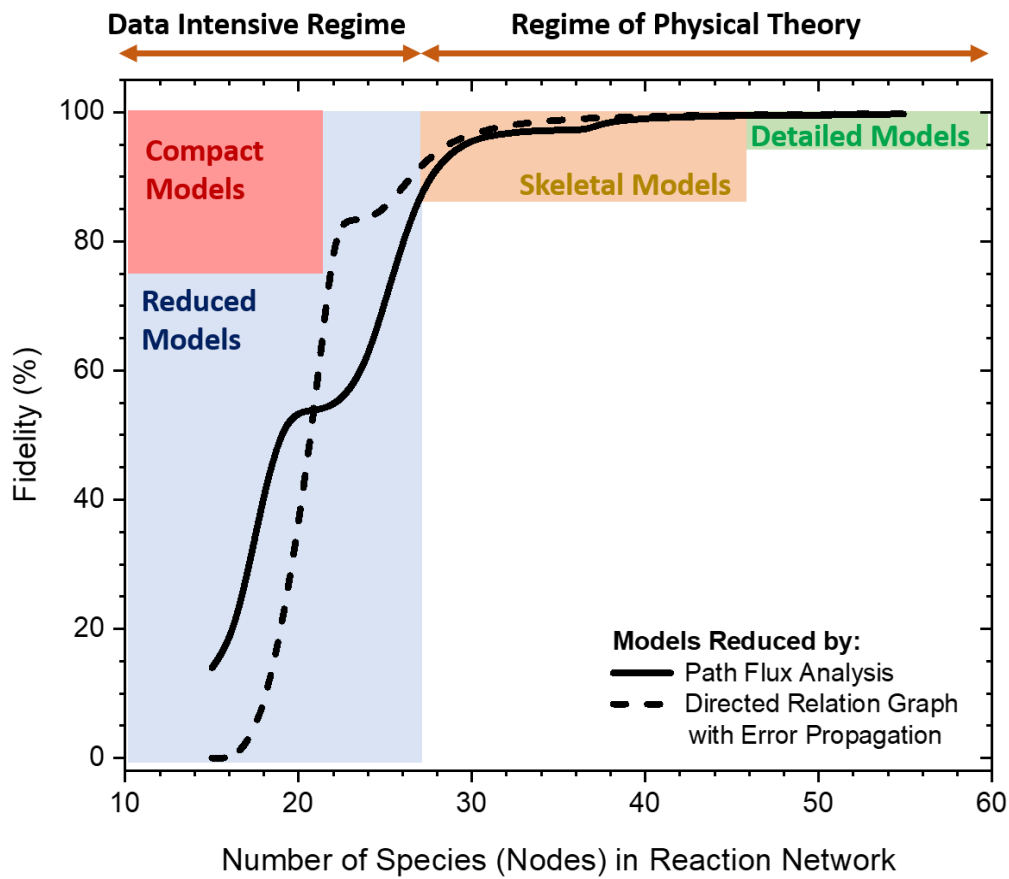

Figure 1. Dependence of fidelity of model calculation for methane combustion on complexity of reaction network, as indicated by number of species (or nodes), where *Detailed, Skeletal, Reduced* and *Compact Models* occupy different regions of detail-fidelity space. A series of skeletal and reduced models are produced by the Directed Relation Graph with Error Propagation reduction method [2] via DoctorSMOKE++ [3, 4], and by the Path Flux Analysis method [6] from the same detailed model [10]. The reduction and fidelity assessment were performed at: 1-40 atm, 1100-2000 K, methane/air mixture fractions 0.5 – 1.5, as described in the text.

The compact model concepts interfaces the fundamental physics-based description of reaction kinetics with a less physical, data-science perspective through the four archetypal kinetic model components:

1. *A Virtual Reaction Network* in place of the reaction mechanism. The purpose of the virtual reaction network is to provide minimal, but sufficient, degrees of freedom to the reacting flux such that certain combinations of virtual reaction rate constants can exist that yield models of high fidelity to the set of reaction kinetic target properties defined by the user. Formally, the virtual reaction network is comprised of nodes (species) joined together by connections (virtual reactions) of varying weights (virtual reaction rate constants). The configuration of each is adjustable to a possibly inexhaustible number of combinations with the objective of replicating some defined set of combustion kinetic performance attributes. The more complex or higher accuracy sought in performance attributes (e.g., species fractions, combustion phenomena), and the wider the range of performance conditions (e.g., mixture fraction, temperature, pressure), the more complex the virtual reaction network needed.

2. *Virtual Reaction Rate Constants (or Weights)* define the magnitude of the connection between the nodes in the virtual reaction network. Analogous to the elementary reaction rate constants, but not describing of a physically real chemical reaction, units are $cm^3mol^{-1}s^{-1}$.

3. *Molecular Thermodynamic Descriptors* including standard state molar enthalpy (H) and entropy (S), and the constant pressure heat capacity ($C_P$) of each node.

4. *Mass and Energy Transport Descriptors* for each node which are used to solve transport equations and diffusion coefficients.

Compact kinetic model production (or *compaction*) involves generating a virtual reaction network, commonly by applying mechanism reduction techniques to a detailed kinetic model, followed by optimisation of the weights and/or network nodes' thermomolecular properties to combustion property targets, such that the resulting compact model is accurate across a defined set of combustion calculations. As neither the thermomolecular properties nor weights are not intended to be physically accurate, their values are not constrained to experimentally measured values.

**1.2.1 Kinetic Model Optimisation**

Kinetic model optimisation (or sometimes *mechanism optimisation*) is the altering of the input parameters of one (or more) of the kinetic model's components to maximise the fidelity to the target set of combustion properties. This most commonly involves optimising the reaction kinetics component through optimisation of the reaction rate constants of important reactions to experimental data or detailed model calculations of important combustion kinetic properties, such as ignition delay time (IDT) and laminar flame speed ($S_u$).

Review of the literature shows that there have been two distinct applications of kinetic model optimisation to kinetic models.

The first is as a method of uncertainty quantification and minimisation for detailed kinetic models. As the Arrhenius reaction rate constant parameters used in detailed kinetic models are obtained through experimental or theoretical calculations, they have a sizeable uncertainty associated with their values. Therefore, when the uncertainty of each reaction rate constant is combined, the resulting uncertainty in the kinetic model is significant. As a result, the model's calculations often fail to replicate, with acceptable accuracy, the results of physical experiments. To fix this, the reaction rate constants are commonly varied within their uncertainty range to find a set of values that results in a kinetic model with a lower overall uncertainty and a better replication of experimental data. Most works of this type use optimisation methodologies that originate from the solution mapping method developed by Frenklach and Miller [25, 26]. This application of mechanism optimisation can be thought of as "fine-tuning", meaning the reaction rate constants do not need to be significantly perturbed to find the ideal set of parameters. This makes the optimisation task relatively straight-forward.

The second application of kinetic model optimisation, and the one of interest to this work, is the optimisation of reduced kinetic models to compensate for the loss in fidelity that is incurred during the reduction process. This most commonly involves adjusting the reaction rate constant parameters to embody the authentic reaction flux through the more limited set of species, such that important aspects of the detailed model's combustion behaviour are reproduced.

There are various methods used to find the ideal set of kinetic parameters for a low-dimensional kinetic model, most commonly genetic algorithm approaches (e.g., [13, 27-34]) and techniques emphasising artificial neural networks (e.g., [35-38]). In these techniques, kinetic models are iteratively produced with new Arrhenius reaction rate constant parameters for selected reactions. These are then evaluated through an *objective error function* (or *fitness function*) which quantifies the error in the new model's calculation of the optimisation targets. The main differences between these techniques are the method of new parameter generation, the degree to which previous solutions influence the performance of the algorithm ("self-learning"), and the required degree of user input ("automation").

The suitability and effectiveness of an optimisation technique depends on the degree of over-reduction of the kinetic model. For example, it was found that during compaction of kinetic models for methane/air combustion, optimisation of a fifteen-node model was many times harder than one with nineteen nodes [21]. The more severe the degree of over reduction, the lower the fidelity of the "base" un-optimised model, and the worse the original Arrhenius reaction rate constant parameters perform as starting points in the optimisation procedure. This means a larger region of the parameter bound space must be searched, and hence, a more complex optimisation technique is required. In other words, the difficulty of the optimisation procedure is coupled to the degree of reduction of the kinetic model. Consequently, an optimisation technique that is successful in the case of a nineteen-node model, may not be successful in the case of one with fifteen nodes.

In compaction, the optimal values of a weight may vary from the original values by orders of magnitude, necessitating a large scan of the weight's bound space. This results in a highly structured error function landscape, containing many ridges and valleys. Due to this complexity, the task of finding the optimum set of parameters is challenging with many optimisation methodologies failing to produce a high-fidelity compact model. Several works on overly-reduced models do not disclose the optimisation method used [7, 16, 17], implying it was performed ad-hoc through manual adjustment of the input parameters. Due to the magnitude of possible solutions and the complexity of the objective error function topology, this approach is severely time-limiting.

In the absence of a rigorous and automated compaction methodology, a significant level of human effort and expertise is needed to produce an accurate compact kinetic model that can be reliably used in multi-dimensional reacting flow simulations. This is a rapidly growing field of research with recent works attempting to improve the process, to yield more accurate low-dimensional ChemKin-form kinetic models with less human involvement [11, 13, 21, 28, 32, 35]. However, a significant advance in this field would be the creation of automated compaction procedure that produces minimally sized, ChemKin-compatible, high-fidelity kinetic models that requires minimal user input. This would enable combustor design engineers to easily obtain compact models that are accurate across the range of phenomena and operating conditions that they require, removing the constraint of user knowledge on the fidelity of the model, allowing for a significant advancement in the efficiency of computational tools in the combustor design process. With the processing power of computers increasing every year, the ability to process large volumes of data may hold the key to developing a method that realises this goal.

In this study, we present a novel compute intensification automated compaction methodology, termed *Machine Learned Optimisation for Chemical Kinetics (MLOCK)*, which is a step towards the goal of a fully automated self-learning compaction methodology. This methodology makes use of modern computing power to perform a large number of computer simulations quickly and relies on these capabilities to learn discreet combinations of kinetic model parameters which result in one or several compact kinetic models which accurately replicate a target set of combustion kinetic calculations. A central tenet of the MLOCK method is that as little prior knowledge as possible be utilised in the compaction procedure. MLOCK thus achieves this through the building of an initial database that results from the intentionally "stupid" sequential perturbation and optimisation of each of the four basic components of the detailed model construct. From this, machine-learning techniques can be applied to deduce the optimal input parameters for each component.

The FAIR [39] principles of data science require that computational models be interoperable, meaning they can be readily used with current software infrastructure ("plug and play"). This has important implications that limit how MLOCK, or other compaction methodologies, should produce compact models. Most commercial combustion modelling platforms allow the ChemKin format of chemical reaction and transport model description. It is thus logical that MLOCK be constructed to maintain this format of reaction description and to operate by perturbation of these parameter descriptors i.e., Arrhenius, NASA polynomial and transport file parameters. This therefore excludes the use of such analytical techniques described earlier, or other techniques that result in a model that does not abide by the ChemKin format and/or requires modification of the solver to be exercised. This paper details the first version of MLOCK, *MLOCK1.0*, which is demonstrated through the application to methane/air combustion. In this work, the newly released NUIGMECH1.0 mechanism [10] is used and a comprehensive account of the MLOCK methodology is presented.

## 2. Methodology – Machine Learned Optimization of Chemical Kinetics (MLOCK)

MLOCK1.0 is a five-step compaction algorithm that assembles a minimalistic virtual reaction network from a detailed reaction network, and subsequently alters the virtual reaction flux through the network by optimising the weights of important connections to a set of target combustion kinetic calculations. The MLOCK algorithm is implemented through python and makes use of the Cantera [40] suite to perform chemical kinetic simulations. Although MLOCK can perturb all four kinetic model components, MLOCK1.0 perturbs only the reaction network and reaction kinetics components.

### 2.1 Perturbation of Reaction Network and Reaction Kinetic Components

The first step in MLOCK1.0 is the assembly of a virtual reaction network by applying path flux analysis (PFA) to remove all non-essential chemical species from a detailed chemical kinetic model's reaction network. Due to the over-reduction of the reaction network, the fidelity of the overly-reduced model's calculations to the validation targets are poor, necessitating a kinetic model optimisation procedure. The reaction kinetics component of the overly-reduced model is then perturbed by optimising the weights of important connections in the virtual reaction network that show an influence on the calculation of the optimisation targets, through simultaneous perturbation of the Arrhenius virtual reaction rate constant parameters ($A$, $n$, and $E_A$), referred to as *virtual Arrhenius parameters* henceforth.

Figure 2 shows the MLOCK optimisation procedure as a series of steps, split across four *Phases*. In each phase, an Objective Error Function (OEF), which quantifies the error in a model's calculation of the selected optimisation targets, is constructed. The overly-reduced model is first optimised to zero-dimensional (0-D) calculations in Phase 1, followed by optimisation to one-dimensional (1-D) calculations in Phase 2. Finally, the model's performance to 0-D and 1-D calculations at the lower and upper pressure limits of the performance envelope are fine-tuned in Phases 3 and 4.

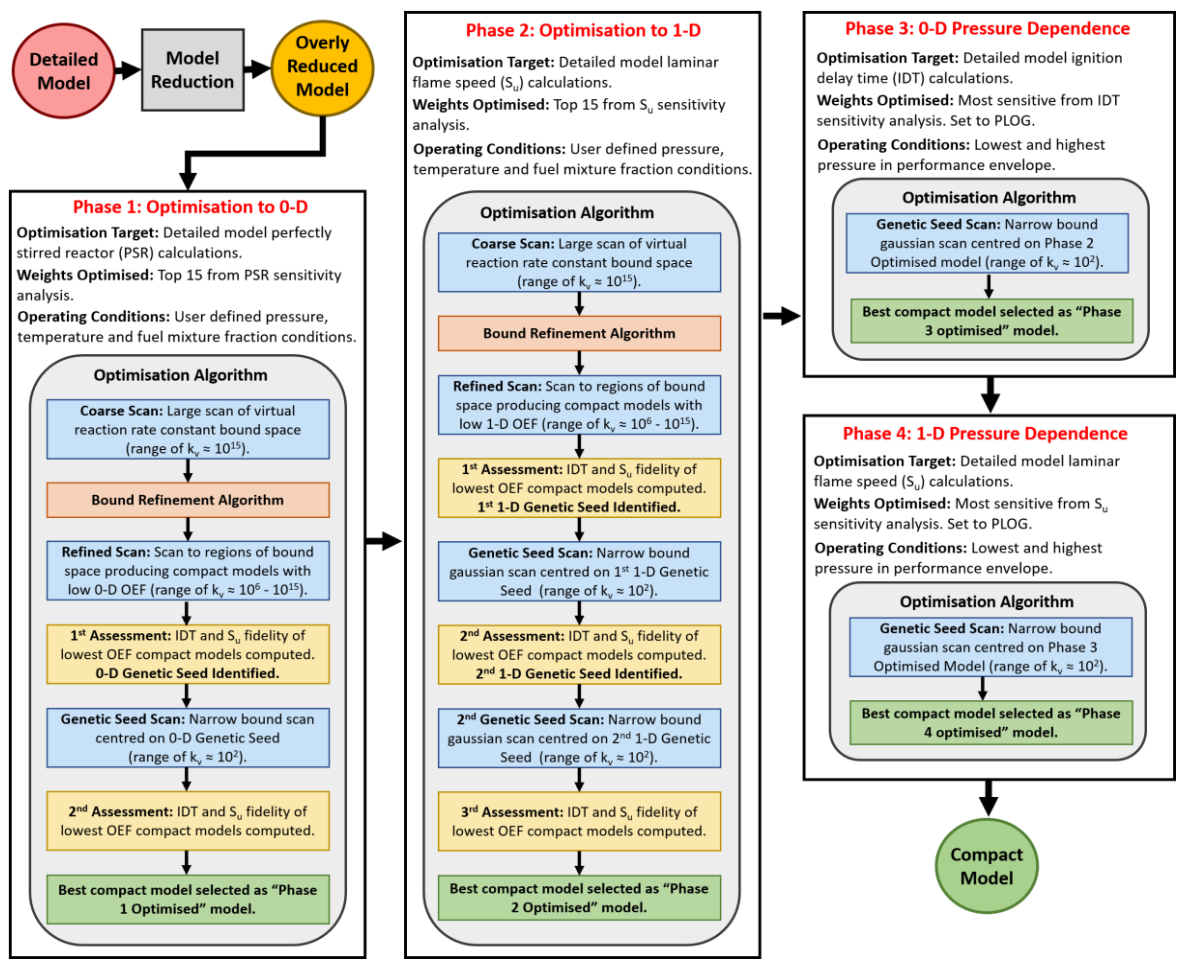

Figure 2. Process flow of the MLOCK1.0 algorithm. For the models discussed in this study, note that (i) the Optimisation Target for Phase 1 is perfectly stirred reactor calculations of OH, H, CO, and $CH_4$ mole fraction profiles at conditions of 1000 – 2000 K, 10 atm and fuel/air equivalence ratios of 0.7 – 1.4. (ii) the Optimisation Target for Phase 2 is freely propagating flame calculations at conditions of 573 K, 10 atm and fuel/air equivalence ratios of 0.8 – 1.5. (iii) Phase 3 Optimisation is performed on constant volume & internal energy calculations (IDT) at conditions of 1100 – 2000 K, 1 and 40 atm and fuel/air equivalence ratios of 0.5 – 1.5. (iv) Phase 4 optimisation is specified to freely propagating flame calculations at conditions of 573 K, 1 and 40 atm and fuel/air equivalence ratios of 0.8 – 1.5. Note that the developmental MLOCK1.0 algorithm and developmental code, allows for the definition of any other set of variables of interest to the user.

Each Phase is composed of one or more *Scans*, which minimise the OEFs through the generation and evaluation of 80,000 compact model candidates by systematic simultaneous perturbation of the virtual reaction rate constants (weights) $k_v$, of the selected connections. This is performed by random combination of each of the *A, n,* and $E_A$ terms of the standard ChemKin format. These scans search an area of the $k_v$ bound space for each connection undergoing optimisation, to assess the suitability of that region in producing high fidelity compact models. The difference between these scans is the magnitude of the $k_v$ bound space being examined, the type of distribution(s) used to produce new virtual Arrhenius parameters, and the OEF being minimised.

Phases 1 and 2 begin with a *Coarse Scan* which generates an approximate representation of the $k_v$ bound space for each connection undergoing optimisation. In this scan, large ranges of $k_v$ are searched for each connection (approximately 15 orders of magnitude), with logarithmically uniform sampling of the $k_v$ range. A *bound refinement algorithm* then uses the information obtained in the Coarse Scan to determine the probability of an accurate compact model candidate being produced in each region of the $k_v$ bound space for each connection.

A *Refined Scan* is subsequently performed with the search algorithm constrained to areas of the $k_v$ bound space identified by the bound refinement algorithm as having a suitably high probability of producing high-fidelity compact models. This increases the resolution of these promising regions of the $k_v$ bound space, increasing the algorithm's efficiency, and leading to the finding of higher fidelity compact model candidates. Following these two scans, an *Assessment* is performed and the best performing compact model candidate is identified and selected as the *Genetic Seed*. The Genetic Seed is considered as possessing favourable characteristics and inhabiting a promising region of the parameter bound space. As such, MLOCK uses this model to perform a narrow search of the parameter bound space using gaussian distributions centred on the genetic seed's virtual Arrhenius parameters. This further directs MLOCK to a region of the bound space capable of producing more accurate compact models. Hence, the efficiency of the optimisation algorithm and fidelity of the compact model improves as it progresses through each Phase.

## 2.2 MLOCK Phase 1

The first optimisation module of MLOCK1.0, *Phase 1*, is optimisation to zero-dimensional (0-D) calculations. A series of detailed model perfectly stirred reactor (PSR) calculations of important chemical species mole fraction profiles are used as the optimisation targets, as they provide a rich source of information on the kinetics of the system. PSR simulations assume a perfectly homogeneous reaction mixture, thereby emphasising a kinetically driven reaction system, where, as in conventional model reduction, an accurate replication of PSR calculations is a good indication that the pertinent kinetics are accurately described. Sensitivity analysis is performed on the overly-reduced model using PSR calculations to identify connections that are important in the model's reactivity at the set of conditions included in the optimisation targets. The sensitivity coefficient of each connection, $\frac{dt_{50\%}}{dk_i}$, is calculated as:

$$\frac{dt_{50\%}}{dk_i} = \frac{|t_{50\%} - t^0_{50\%}|}{t^0_{50\%} * c} \qquad (4)$$

Where $t_{50\%}$ is the time for 50 % fuel consumption in a PSR following perturbation of a weight by a factor $c$ and $t^0_{50\%}$ is the time for 50 % fuel consumption prior to perturbation. The fifteen connections with the largest sensitivity coefficient are selected for optimisation.

As the performance of a compact model candidate is defined by the magnitude of the OEF, the selection of properties to evaluate and the mathematical form of the OEF is crucial. As we are primarily interested in the performance of the compact model in replicating the detailed model calculations at times close to ignition, an OEF is required that prioritises the error in the calculations at this time. Equation 5 shows the OEF used in Phase 1 (*0-D OEF*). The first term in this OEF is designed such that the error is weighted more heavily at times when the mole fraction of the specific species is close to the maximum mole fraction. This term was used by Nagy et al. [41] to evaluate the error in species concentrations during model reduction. However, as the fuel always has a maximum mole fraction at the start of the PSR simulation (t = 0), this term does not provide a suitable evaluation of its error. As such, if the mole fractions of the fuel are included as an optimisation target, the error is evaluated by a simple mean squared error.

$$\text{0-D OEF} = \frac{1}{N_J N_K} \sum_k \sum_j \left\{ \left[ \sum_i w_i \left( 2 \frac{|X^{red}_{i,k}(t_j) - X^{det}_{i,k}(t_j)|}{X^{det}_{i,k}(t_j) + max(X^{det}_{i,k})} \right) \right] + w_{fuel} \left( X^{red}_{fuel,k}(t_j) - X^{det}_{fuel,k}(t_j) \right)^2 \right\} \qquad (5)$$

Where, $X_{i,k}$ refers to the mole fraction of the $i'^{th}$ species at the $k'^{th}$ set of temperature, pressure, and equivalence ratio conditions. $N_K$ and $N_J$ refer to the total number of operating conditions and time points that are included in the evaluation, and *w* is a weighting term that can be used to assign a greater importance to the error in certain species mole fractions.

## 2.3 MLOCK Phase 2

Following Phase 1, the best performing compact model candidate is identified and subsequently optimised to the detailed model's laminar flame speed ($S_u$) calculations, which act as an indicator for model performance at 1-D calculations. $S_u$ sensitivity analysis is performed on the Phase 1 Optimised model, and the top fifteen most sensitive connections at the set of conditions included in the optimisation targets are selected for optimisation. As in Phase 1, Phase 2 employs the Coarse Scan – Bound Refinement – Refined Scan – Genetic Seed Scan procedure, however, an additional genetic seed scan is performed with the optimisation targets comprised of laminar flame speed calculations at rich conditions. Despite the compact model candidates replicating the laminar flame speed well at fuel-lean conditions following the *1st Genetic Seed Scan*, the fidelity at rich conditions is usually low, necessitating an additional optimisation step at rich equivalence ratios. This decrease in fidelity at fuel-rich conditions is a common characteristic of compact models as they do not contain the necessary hydrocarbon species that are crucial in providing additional chemical pathways for the reaction flux to result in an accurate description of the more complex fuel-rich chemistry. The error in a compact model's calculation of the optimisation targets is evaluated through the Phase 2 OEF (1-D OEF):

$$1\text{-}D \ OEF = \frac{1}{N_k} \sum_k \left( \frac{|S_{u,k}^{red} - S_{u,k}^{det}|}{S_{u,k}^{det}} \right) \quad (6)$$

Where $S_{u,k}$ is the laminar flame speed at the $k'^{th}$ set of temperature, pressure, and equivalence ratio conditions and $N_k$ is the number of conditions. The best performing model resulting from this second phase of optimisation is identified as the "Phase 2 Optimised" model.

## 2.4 Bound Refinement

The Coarse Scan in Phases 1 and 2 samples $k_v$ across a range of approximately fifteen orders of magnitude to identify a relationship between the value of $k_v$ and the performance of the compact model, for each connection. The degree of coupling between the value of $k_v$ for a connection and the OEF varies across the set of connections undergoing optimisation. For some connections the coupling is strong, with the existence of a clear global minimum, however for others the coupling is more complex with the existence of multiple minima (See Supplemental Material). The more connections you optimise, the less clear the minima are, i.e., an increase in the levels of "noise" in the data and "broadening" of the minima.

The goal is to use the information from the Coarse Scan to improve both the efficiency and effectiveness of the optimisation algorithm. This is done by examining the topography of the OEF landscape as a function of $k_v$ for each connection and constraining the search algorithm to regions of the parameter space that produce compact model candidates with a low OEF. This can be done manually, however this requires considerable human input. Therefore, a bound refinement algorithm was created to automatically identify these promising regions of the $k_v$ bound space. It does this by using the information from the Coarse Scan report to score each region of the $k_v$ bound space on the likelihood of a good compact model candidate being produced in that region. This algorithm works on the principal of a binomial test in which the null hypothesis is that the percentage of compact models in a region of $k_v$ that have an OEF below a certain value, is the same as the percentage across the whole population (full $k_v$ range). A Z-test is then performed to test if the null hypothesis is true or false for each region of $k_v$.

First, the value of the OEF that corresponds to the $x$'th percentile of the 80,000 compact model candidates produced in the Coarse Scan is calculated and the number of compact model candidates in the complete population with an OEF below this value is recorded. The $k_v$ bound space is then divided into a series of regions, and the number of compact models in each region that have an OEF below this threshold is counted. Each region is then scored based on the percentage of compact model candidates in the region with an OEF below this threshold. If the proportion of models below the OEF threshold in the full population is $\bar{p}$, and within a given region of $k_v$ it is found to be $p$, then for a region of $k_v$ with $n$ compact model candidates, the Z-score for that region is,

$$Z = \frac{\bar{p} - p}{\sqrt{\frac{\bar{p}(1 - \bar{p})}{n}}} \tag{7}$$

Lower Z-values correspond to a greater proportion of low OEF compact model candidates produced in the specific $k_v$ region. A significance (α) value of 0.05 is set and the algorithm outputs regions of $k_v$ with a Z-score below the critical value, implying a larger proportion of "good" models in the sample than the population. This information is subsequently used by the Refined Scan to constrain the search algorithm to areas of $k_v$ bound space identified by the bound refinement algorithm as producing a large proportion of high-fidelity models This is performed independently for every connection undergoing optimisation.

An example of this process for the connection $CH_3 + O_2 \Leftrightarrow CH_3O + OH$ during the compaction of a fifteen-node methane/air kinetic model is shown in Figure 3. Compact model candidates produced in the Coarse Scan (black symbols) and in the Refined Scan (red symbols) in Phase 1 are plotted as a function of their 0-D OEF, with the starting unoptimised overly-reduced model represented by a pink star. The bound refinement algorithm identifies the region in the $k_v$ bound space between approximately $10^8$ and $10^{12}$ as having a higher probability of producing better models (blue line). Therefore, only compact model candidates with a value of $k_v$ at 2000 K for this connection between $10^8 - 10^{12}$ are produced in the Refined Scan. Following the Refined Scan there exists a group of models with a similar 0-D OEF, however, it is likely that a variation exists in the model's fidelity to other validation targets. Therefore, to identify the best overall compact model, the IDT and $S_u$ fidelity of the compact model candidates with the lowest 0-D OEF are calculated in the *Assessment* step. The compact model candidate that meets the defined selection criteria is selected as the Genetic Seed (blue triangle). A Genetic Seed Scan is then performed, centred on this best model's parameters, further populating the region of $k_v \sim 10^{10}$.

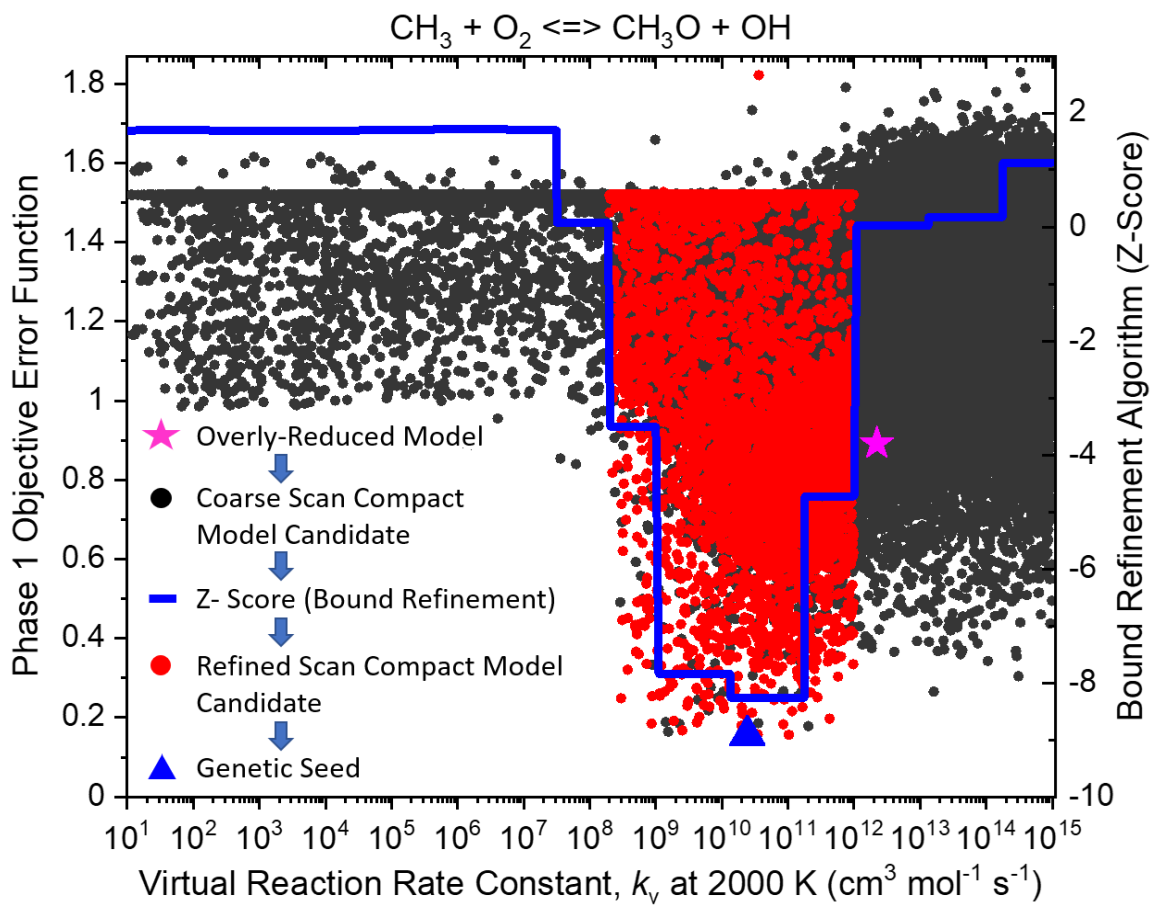

Figure 3. Phase 1 objective error function (OEF) of compact model candidates produced in the Phase 1 coarse and Refined Scans as a function of the virtual reaction rate constant of the connection $CH_3 + O_2 <=> CH_2O + H$. The Z-score is calculated by the bound refinement algorithm following the Coarse Scan and directs the search algorithm to regions of the virtual reaction rate constant bound space that produce models with low OEF. The unoptimised model is that produced by path flux analysis. 2000 K is used as a nominal reference.

## 2.5 MLOCK Phase 3 and Phase 4

Within the ChemKin framework, there exists a standard reaction expression termed *PLOG,* which is exploited in this work to provide additional degrees of freedom. This allows for the explicit expression of a set of Arrhenius parameters for discreet pressures. If the pressure during a calculation is between the pressure points provided in the PLOG expression, the reaction rate is obtained by logarithmic interpolation of the specified Arrhenius parameters. Therefore, the pressure dependence of a reaction rate is based on direct interpolation of reaction rates at specified pressures. For a pressure $P$, between $P_i$ and $P_{i+1}$, the reaction rate constant $k_v$ is calculated by;

$$\ln k = \ln k_i + (\ln k_{i+1} - \ln k_i) \frac{\ln P - \ln P_i}{\ln P_{i+1} - \ln P_i} \qquad (8)$$

Thus, in this work, a new set of Arrhenius parameters can be defined and optimised for various pressures to fine-tune the performance of the compact model at these pressures, without affecting the calculations of the compact model at a previously optimised pressure(s). As the virtual reaction network of a compact model is not intended to be physically authentic, it is valid to impose pressure dependence on a reaction that does not depend on pressure in reality. Therefore, the IDT fidelity of the Phase 2 Optimised model is calculated at the lowest and highest pressure within the performance envelope. If the compact model does not perform to a satisfactory standard at either pressure, it is optimised to IDT calculations at that condition to tune the performance of the model to capture this dependence on pressure. IDT sensitivity analysis is performed at the specified pressure and the most sensitive connection is identified and set to PLOG. As Phases 3 and 4 are fine-tuning optimisation modules, only one connection is optimised. The original Arrhenius parameters for the connection are set as the expression for the pressure(s) used in Phases 1 and 2. A Genetic Seed Scan is then performed centred on the original Arrhenius parameters. Following this, the "Phase 3 Optimised" model enters Phase 4, where the process is repeated but for the case of laminar flame speed. The best performing compact model candidate from Phase 4 is identified as the final *Compact Model*.

## 3. Results with Application to Methane/Air Combustion

MLOCK1.0 was applied to the NUIGMECH1.0 detailed chemical kinetic model containing 2,759 species, to produce a compact kinetic model for methane/air combustion, containing 15 nodes in the virtual reaction network. To wholly quantify model accuracies, Fidelity Indexes of the form of Equation 9 are defined to simply sum the relative errors of each point in the overall performance set,

$$Fidelity\ (\%) = \frac{100}{N} \sum_k \left( 1 - \frac{|X_k^{compact} - X_k^{detailed}|}{X_k^{detailed}} \right) \quad (9)$$

where $X_k$ is the calculation (e.g., IDT) at the $k^{th}$ set of pressure, temperature, and equivalence ratio conditions and $N$ is the number of conditions included.

PFA was performed on NUIGMECH1.0 to construct a virtual reaction network containing a minimal number of chemical species. A numerical database of time dependent chemical species concentration profiles in a homogenous constant volume reactor up until the point of ignition was populated with simulations using methane/air mixtures at 1, 10, 20, 30, and 40 atmospheres (atm), 900 – 2000 K and at equivalence ratios 0.5, 1 and 1.5. To examine the extinction phenomenon, a second numerical database populated at conditions of 1, 10, 20, 30 and 40 atm, 300 - 1500 K and equivalence ratios 0.5, 1 and 1.5 was constructed using perfectly stirred reactor (PSR) simulations. The resulting overly-reduced model was tested against PSR, IDT and $S_u$ calculations at 10 atm to assess its performance. The model had a PSR fidelity across 1000 – 2000 K of 38.3%, an IDT fidelity across 1100 – 2000 K of 16.9% and a $S_u$ fidelity at 573 K of 58.6%.

The performance target parameter space is vast, and so only representative calculations can be shown here with the intention of illustrating the behaviour of MLOCK. Full accounting of the resulting compact model is provided in the supplemental material. An important note is that MLOCK1.0 was developed using methane which is a well-researched fuel and as such there exists detailed kinetic models that replicate the reaction mechanism of methane combustion and the resulting combustion phenomena to excellent fidelity. In the case of less researched fuels where an accurate detailed model is not available, MLOCK1.0 can be easily adapted to use experimental measurements as the optimisation targets.

Table 1. Optimisation targets and validation targets (used in the Assessment steps) used by MLOCK1.0 in this study to provide an indicator for model performance in 0-D and 1-D calculations at Industry relevant operating conditions.

| Purpose | Reactor Physics | Property | Pressure (atm) | Phi | Temperature (K) |
|---|---|---|---|---|---|
| **Phase 1 OEF** | Perfectly Stirred | [$CH_4$], [CO], [OH], [H] | 10 | 0.7, 0.9, 1.0, 1.1, 1.4 | 1000, 1500, 2000 |
| **Phase 1 Validation** | Ignition Delay Time (Constant UV) | IDT | 10 | 0.5, 1.0, 1.5 | 1100, 1200, …, 2000 |
| | Freely Propagating Flame | $S_u$ | 10 | 0.4, 0.5,…, 1.5 | 573 |
| **Phase 2 OEF** | Freely Propagating Flame | $S_u$ | 10 | 0.8, 1.05, 1.3, 1.5 | 573 |
| **Phase 2 Validation** | Ignition Delay Time (Constant UV) | IDT | 10 | 0.5, 1.0, 1.5 | 1100, 1200, …, 2000 |
| | Freely Propagating Flame | $S_u$ | 10 | 0.4, 0.5,…, 1.5 | 573 |
| **Phase 3 OEF** | Ignition Delay Time (Constant UV) | IDT | 1, 40 | 0.5, 1.0, 1.5 | 1100, 1200, …, 2000 |
| **Phase 4 OEF** | Freely Propagating Flame | $S_u$ | 1, 40 | 0.8, 1.05, 1.3 | 573 |
| **Phase 4 Validation** | Ignition Delay Time (Constant UV) | IDT | 1, 40 | 0.5, 1.0, 1.5 | 1100, 1200, …, 2000 |
| | Freely Propagating Flame | $S_u$ | 1, 40 | 0.4, 0.5,…, 1.5 | 573 |

## 3.1 Results of MLOCK Phase 1

Mole fraction profiles of CO, OH, H, and $CH_4$ in NUIGMECH1.0 PSR calculations at 10 atm; 1000, 1500, and 2000 K; and equivalence ratios 0.7, 0.9, 1.0, 1.1, and 1.4, were selected as the optimisation targets to comprise the 0-D OEF in Phase 1, as shown in Table 1. These chemical species were selected on the hypothesis that as they are indictors for fuel decomposition, heat release rate, and the reactivity of the system, their accurate replication is essential in replicating the pertinent kinetics of the detailed model. Figure 4 shows the IDT fidelity of compact model candidates produced in Phase 1 Refined Scan as a function of their 0-D OEF. This shows that the lower a compact model's 0-D OEF, i.e., the better the model replicates the detailed model's PSR calculations for these species, the more likely it is to have a good replication of the detailed model's ignition delay time calculations, confirming the original hypothesis. Similar to previously reported [34], it was found that multiple sets of solutions produced a similarly high-fidelity replication of the optimisation targets, however, fidelity to other validation calculations varied considerably. This is partly observed in Figure 4 where a group of compact model candidates with an OEF of ~0.15, implying good replication of detailed model's PSR calculations, have a noticeable variation in IDT fidelity. Choosing the genetic seed solely based on 0-D OEF would not guarantee the best overall model, hence, the *Assessment* steps are crucial in the efficiency of the algorithm through identifying the best overall compact model candidate, taking into account other performance targets.

In the *Assessment* step, the following procedure was followed to identify the best performing compact model candidate.

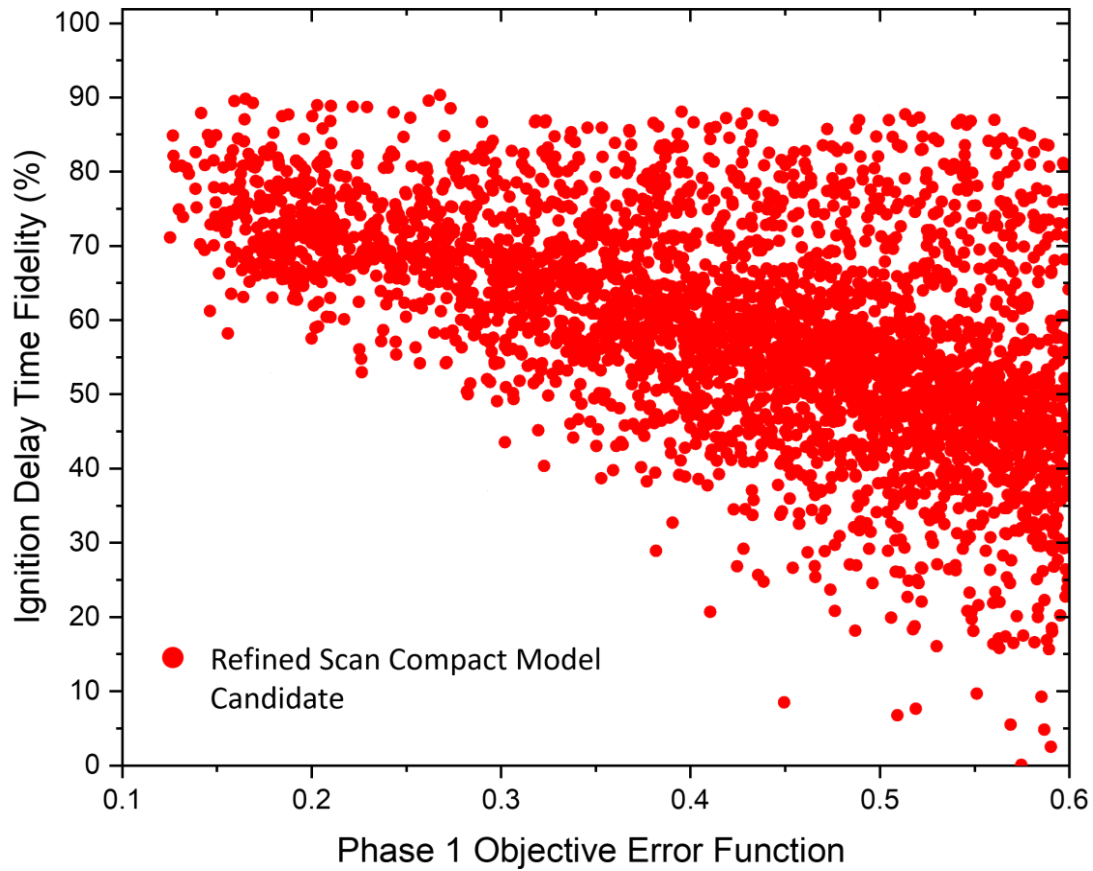

Figure 4. Ignition delay time (IDT) fidelity of 15 species compact models produced in MLOCK Phase 1 Refined Scan as a function of their Phase 1 objective error function (0-D OEF). IDT fidelity was calculated at 10 atm, 1100 – 2000 K, and equivalence ratios 0.5 – 1.5.

1. IDT and $S_u$ fidelity of the 150 compact model candidates with the lowest OEF is calculated at the range of conditions shown in Table 1.
2. Model candidates that do not experience a maximum flame speed at $\phi = 1$ or 1.1 are removed from the database.
3. Model candidates that experience a maximum flame speed greater than $\pm 25\%$ of NUIGMECH1.0 are removed from the database.
4. The compact model candidate with the largest average $S_u$ and IDT fidelity is selected as the best compact model candidate.

These selection criteria are applied to avoid selecting a compact model with an un-realistic flame speed behaviour and to obtain the best overall model. This selection process was applied at all Assessment steps.

Figure 5 shows the number of compact model candidates produced in the Phase 1 Coarse, Refined, and Genetic Seed Scans that have a Phase 1 OEF less than 0.3 (i.e., a good replication of NUIGMECH1.0 PSR calculations). This shows the improvement in the efficiency of MLOCK as it proceeded through Phase 1. Following Phase 1, the PSR, IDT, and $S_u$ fidelity of the compact model improved from 39.3% to 89.8%, 16.9% to 80.4% and 55.1% to 79.1% respectively across the Phase 1 validation calculations shown in Table 1.

### 3.2 Results of MLOCK Phase 2

The optimisation targets used in Phase 2 consisted of NUIGMECH1.0 calculations of laminar flame speed at 573K, 10 atm, and at equivalence ratios 0.8, 1.05 and 1.3. As the computational cost in simulating a flame is much greater than that of a PSR due to the added dimensionality necessitating the consideration of the effect of species transport, the number of calculations included in the 1-D OEF is kept at a minimum to balance the computational efficiency with the effectiveness of the algorithm. Including three targets in the 1-D OEF was found to be a suitable compromise between computational cost and confidence in the OEF as an indicator for performance across the complete validation calculation space. Following the *1$^{st}$ Genetic Seed Scan*, the best performing model is selected as the *2$^{nd}$ 1-D Genetic Seed* and optimised to the $S_u$ at equivalence ratios 1.05, 1.3, and 1.5 to improve the performance of the model at fuel-rich conditions. Despite the performance of the compact model at fuel-rich conditions improving following the *2$^{nd}$ Genetic Seed Scan*, the fidelity of the model is still lower here than at lean and stoichiometric conditions due to the absence of some important pathways from the virtual reaction network e.g., ethane ($C_2H_6$).

The criterion for selecting the Genetic Seed and Phase 2 Optimised model is the same as that used in Phase 1. As a result of Phase 2, the PSR, IDT, and $S_u$ fidelity of the compact model improved from 89.8% to 87.9%, 80.4% to 84.9% and 79.1% to 89.6% respectively across the Phase 2 validation calculations shown in Table 1. Figure 6 shows the result of Phase 2 optimisation where the IDT as a function of temperature at a representative condition, and $S_u$ as a function of equivalence ratio of the unoptimised, Phase 1 Optimised, and Phase 2 Optimised models is plotted.

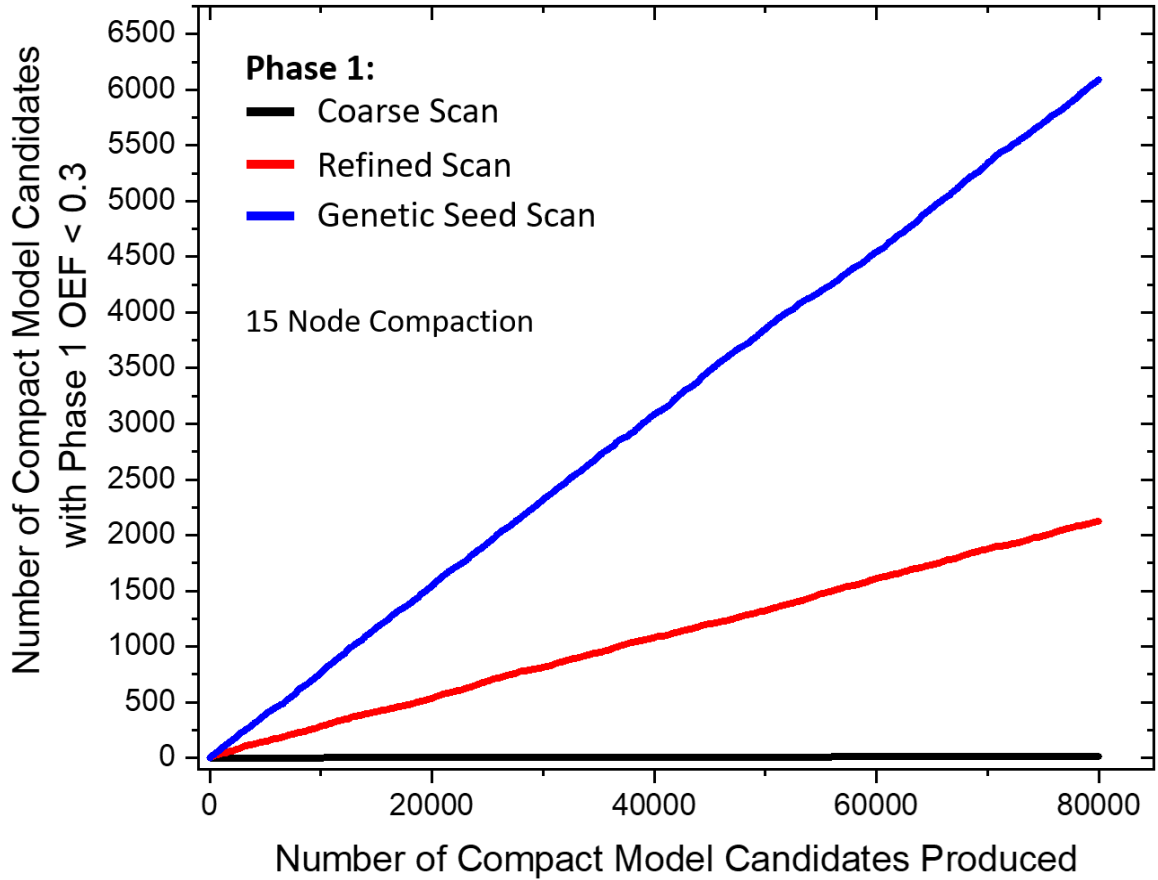

Figure 6. The number of compact models produced with a 0-D OEF less than 0.3 in each of the MLOCK Phase 1 scans as a function of the number of compact models produced.

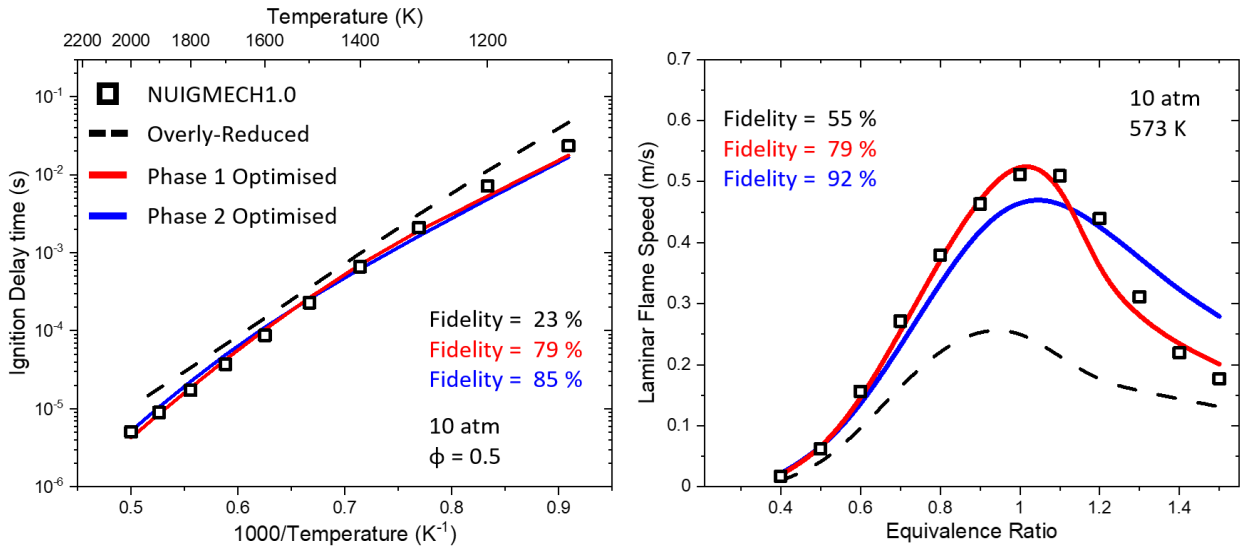

Figure 5. Ignition delay time as a function of temperature (left) and laminar flame speed as a function of equivalence ratio (right) for a methane/air mixture at 10 atm using the NUIGMECH1.0 detailed model (symbols) and the 15 species compact models (line).

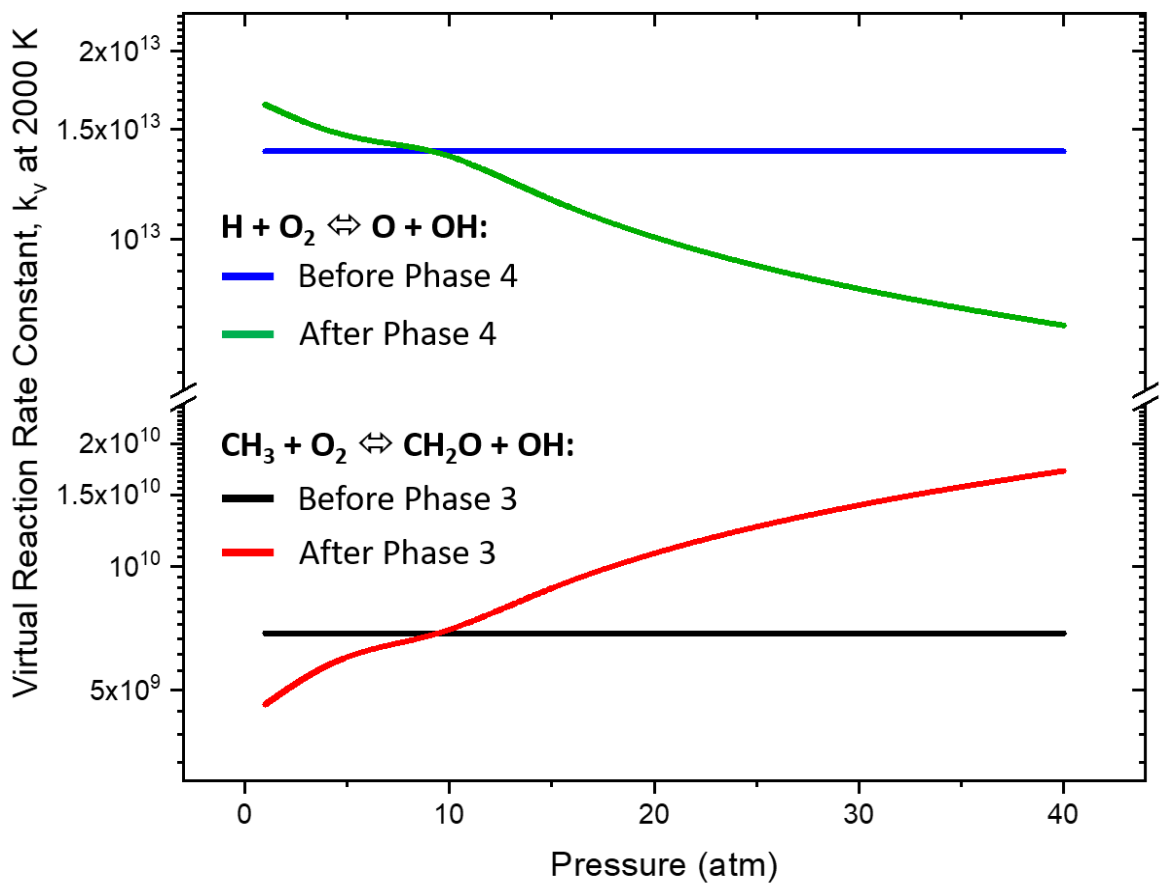

Figure 7. Effect of Phase 3 and Phase 4 on the virtual reaction rate constant for the connections $CH_3 + O_2 \Leftrightarrow CH_2O + OH$ and $H + O_2 \Leftrightarrow O + OH$ respectively.

### 3.3 Results of MLOCK Phases 3 and 4

Phases 3 and 4 used NUIGMECH1.0 calculations of the IDT and $S_u$ respectively at 1 and 40 atm as the optimisation targets to fine-tune the performance of the compact model at these pressures. The most sensitive connection to IDT at 1 atm was set to a PLOG reaction and the original Arrhenius parameters set to correspond to the reaction rate expression at 10 atm. The virtual Arrhenius parameters corresponding to 1 atm were then optimised to NUIGMECH1.0 IDT calculations at 1 atm. This was repeated for 40 atm and the best model selected as the "Phase 3 Optimised" model. The most sensitive connection at both 1 and 40 atm was $CH_3 + O_2 \Leftrightarrow CH_2O + OH$. The value of $k_v$ for this connection was decreased at 1 atm by ~33 % from the optimised value at 10 atm, whereas at 40 atm it was increased by ~ 149 %, as shown in Figure 7. In other words, to improve replication of NUIGMECH ignition delay time calculations at 1 atm, it was found that this connection needed to be slowed down, to reduce the production of the OH radical, whereas at 40 atm the connection must be sped up to increase reactivity. The same procedure was then performed for $S_u$ in Phase 4 on the connection H + O$_2$ ⇔ O + OH. Here, it was found that to improve fidelity to NUIGMECH flame speed calculations at 1 atm, $k_v$ must be increased by ~19% from the optimised value corresponding to 10 atm, whereas at 40 atm it must be decreased by ~ 47%.

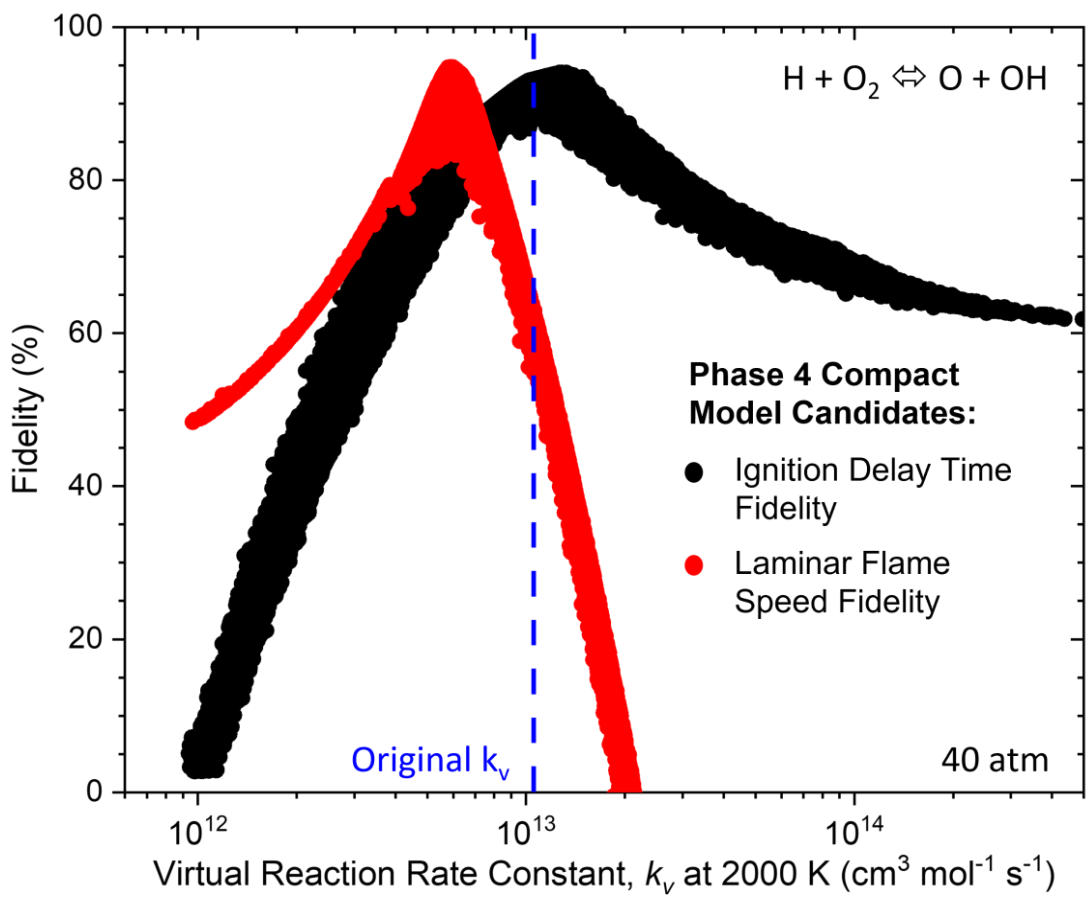

Figure 8. The ignition delay time (black) and laminar flame speed (red) fidelity of 15 node compact model candidates produced during MLOCK Phase 4 at 40 atm as a function of $k_v$ at 2000 K for the connection H + O$_2$ ⇔ O + OH.

Following Phase 4, it was observed that the compact model candidates with the highest IDT fidelity tended to have a low $S_u$ fidelity. This is due to a difference in the value of $k_v$ that corresponds to the global maxima for the $S_u$ and IDT fidelity, as shown in Figure 8, thus necessitating a compromise. The best performing compact model resulting from Phase 4 is then selected as the final Compact Model.

The result of the optimisation process on the compact model's fidelity to the validation calculations at a range of pressures is shown in Table 2.

### 3.4 Comparison to State-of-Art

Table 3 compares the performance of the 15-node compact model to the state-of-art compact models for methane/air combustion across a wide range of conditions. Here, state-of-art is defined in terms of the number of nodes (species) in the virtual reaction network and the accuracy of the compact model's calculations to the parent detailed model. DRM22 is shown as it is a familiar and widely-used reduced model and as such its performance serves as a good benchmarking metric.

The compact model produced in this work outperforms the state-of-art compact models across these broad range of gas turbine relevant conditions and combustion kinetic calculations. However, it is important to note that only the compact models produced by TCD were created with the purpose of describing combustion phenomena across a comprehensive range of operating conditions. Due to the lack of detail in the virtual reaction network of compact models, the task of producing a compact model that is valid across a wide range of conditions is challenging and demands a more complex optimisation technique. Thus, compact models tend to be produced for use in limited combustion calculations at a narrow range of operating conditions. The compact model developed by Lytras et al., for example, was constructed for application to 1-D calculations at atmospheric conditions and so, as expected, the compact model performs well at such calculations (87.0%), however, performs poorly at 0-D calculations and at elevated pressures.

Table 2. Ignition delay time and laminar flame speed fidelity of 15 node model at various stages during the MLOCK process.

| Combustion Property | Model | 1 atm | 10atm | 20 atm | 30 atm | 40 atm |
|---|---|---|---|---|---|---|
| **Ignition Delay Time (IDT)** $\phi = 0.5 - 1.5$ 1100 – 2000 K | Overly-Reduced | 23.2 % | 51.0 % | 48.0 % | 41.4 % | 34.2 % |
| | Phase 2 Optimised | 79.8 % | 84.0 % | 84.7 % | 80.9 % | 75.2 % |
| | Compact Model | 89.6 % | 84.0 % | 87.3 % | 87.6 % | 87.0 % |
| **Laminar Flame Speed ($S_u$)** $\phi = 0.5 - 1.5$ 573 K | Overly-Reduced | 64.4 % | 55.8 % | 57.8 % | 59.8% | 60.9% |
| | Phase 2 Optimised | 67.7 % | 91.9 % | 84.8 % | 78.8 % | 74.9 % |
| | Compact Model | 83.9 % | 91.9 % | 91.9 % | 91.1 % | 90.2 % |

A final important point is regarding the accuracy of the parent detailed models. The fidelities of the compact models in Table 3 are calculated with respect to their parent detailed model. This study utilizes the latest NUIG detailed chemical kinetic model from 2020. It is expected to be more accurate to the reality of methane combustion than detailed models which precede it.

## 4. Discussion

### 4.1 Objective Error Functions

While a strong correlation between the Phase 1 OEF and the IDT fidelity of compact model candidates exists (Figure 4), there is no meaningful correlation between the Phase 1 OEF and the $S_u$ fidelity of compact model candidates, as shown in Figure 9 for models produced in produced in the Phase 1 Refined Scan, necessitating a direct optimisation to flame calculations. This is not a problem when dealing with compact models for simple fuels such as methane, as direct optimisation to laminar flame speed calculations can be afforded. However, this non-correlation can pose a problem when working with more complex systems e.g., liquid fuels. There, flame speed calculations have a considerable computational cost associated with them and can be numerically "stiff" to converge. It is therefore highly desirable to obtain an OEF that is comprised of information from computationally cheap 0-D calculations that provide a good indication of a compact model candidate's performance at both 0-D and 1-D calculations. Achieving this would remove, or significantly lessen, the need for a large number of computationally expensive or numerically difficult 1-D calculations.

Table 3. Comparison of state of art models for methane/air combustion to respective detailed model calculations. IDT fidelity was calculated at 1 – 40 atm, 1100 – 2000 K and equivalence ratios 0.5-1.5; $S_u$ fidelity was calculated across 1 – 40 atm, 473 – 673 K and equivalence ratios 0.4 – 1.5; and PSR fidelity was calculated at 10 atm, 1000 – 2000 K and equivalence ratios 0.7 – 1.4 for error in $CO, CO_2$, OH and $H_2O$ maximum and equilibrium mole fractions.

| Author | No. of Nodes | IDT Fidelity (%) | $S_u$ Fidelity (%) | PSR Fidelity (%) | Parent Mechanism |
|---|---|---|---|---|---|
| *Benchmarking Model* | | | | | |
| DRM22 [1] | 24 | 98.1 | 94.5 | 98.1 | GRI 1.2 [5] |
| *State-of-art Compact Models* | | | | | |
| Lytras et al. [7] | 14 | 18.6 | 65.4 | 81.8 | USC Mech II [9] |
| Leylegian [11] | 15 | 39.3 | 83.6 | 74.9 | USC Mech II [9] |
| Leylegian et al. [13] | 15 | 16.3 | 85.9 | 93.0 | USC Mech II [9] |
| Bioche et al. [14] | 17 | 94.7 | 65.1 | 94.4 | GRI 1.2 [5] |
| Larson et al. [16, 17] | 18 | 0.0 | 81.7 | 10.1 | GRI 3.0 [18] |
| *MLOCK Compact Models* | | | | | |
| This Work | 15 | 87.1 | 88.8 | 87.2 | NUIGMECH1.0 [10] |
| Kelly et al. [21] | 15 | 75.0 | 83.4 | 93.9 | NUIG18_17_C3 [22] |
| Kelly et al. [21] | 19 | 93.6 | 94.9 | 95.1 | NUIG18_17_C3 [22] |

Following review of the literature in this regard, given the prevalence of the use of 0-D computed data in model reduction, it is surprising that apparently no effective relationships to this purpose have been demonstrated. Sikalo et al [42] optimised a 25 species model for methane combustion to an OEF composed of 0-D calculations, leading to a reduction in the error in laminar flame speed calculations from 10% to 5%. Error in the mole fractions of atomic hydrogen in 0-D reactor was included in the OEF as an indicator for flame speed performance due to its importance in the flame. The OEF that was used by Sikalo et al was applied to this work, however no discernible relationship was found between this OEF and the flame speed fidelity for fifteen-node compact model candidates. The likely reason for this is that in a fifteen-node reaction network construct, atomic hydrogen will no longer embody the same characteristics as in a reaction mechanism, and therefore it will not perform as well as an indicator for flame speed performance.

Thus, as an interim measure to bridge this divide, a basic hybrid OEF composed of data from a library of 0-D calculations trained by multiple linear regression to calculations of $S_u$, was devised.

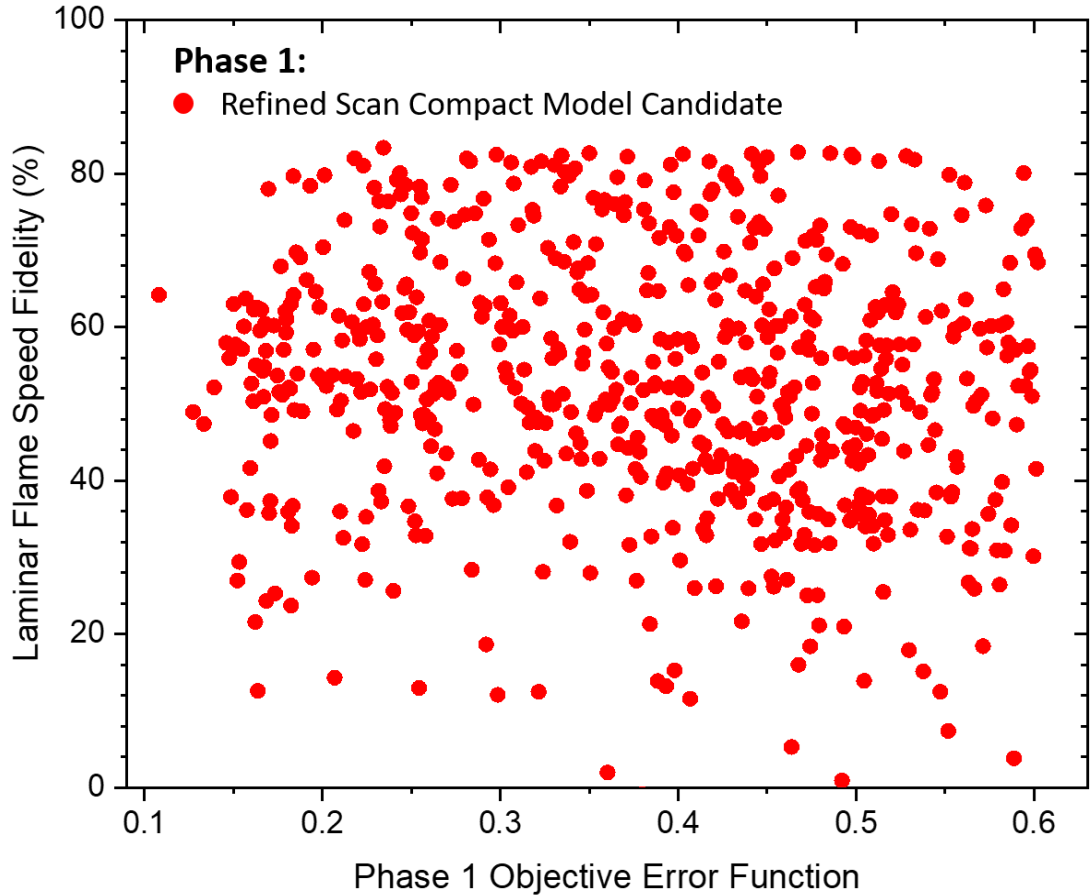

Figure 9. Laminar flame speed fidelity of a subset of 15 species compact model candidates produced by the Phase 1 Refined Scan as a function of the Phase 1 objective error function (OEF). Laminar flame speed fidelity calculated at 573 K, 10 atm, and equivalence ratios 0.4 – 1.5. No discernible dependence of flame speed fidelity to Phase 1 objective error function is apparent.

## 4.2 Multiple Linear Regression

Multiple linear regression is a statistical technique that uses several explanatory variables, $x_i$, to predict the outcome of a response variable, $Y_i$. This can be written as:

$$Y_i = \beta_0 + \beta_1 x_{i1} + \beta_2 x_{i2} + \cdots + \beta_n x_{in} \qquad [4]$$

Where β denotes the regression coefficients for each explanatory variable. Multiple linear regression was employed in Phase 2 to improve the efficiency of the MLOCK algorithm by predicting the flame speed at each condition in the OEF using a series of PSR properties as the explanatory variables. From this, the 1-D OEF was then calculated. A correlation matrix was used to select the properties to include as explanatory variables for each flame speed. The properties are:

(i) Relative error in the maximum mole fractions of selected species,

(ii) Relative error in the steady-state mole fractions of selected species,

(iii) Relative error in the maximum and steady-state temperature,

(iv) Relative error in the maximum heat release rate,

at a range of initial temperatures and equivalence ratios.

The multiple linear regression model was trained and validated on Phase 2 Refined Scan data, and then employed in the Phase 2 Genetic Seed Scans. This enabled the prediction of the Phase 2 OEF for compact model candidates produced in the Phase 2 Genetic Seed scans using a mathematical relationship rather than the simulation of a series of laminar flames for each compact model candidate, significantly reducing the computational cost of MLOCK. Figure 10 shows the predictive capability of the regression model for compact model candidates produced in the 1$^{st}$ Genetic Seed scan in Phase 2. The application of multiple linear regression to the MLOCK algorithm will be further investigated and improved in future work.

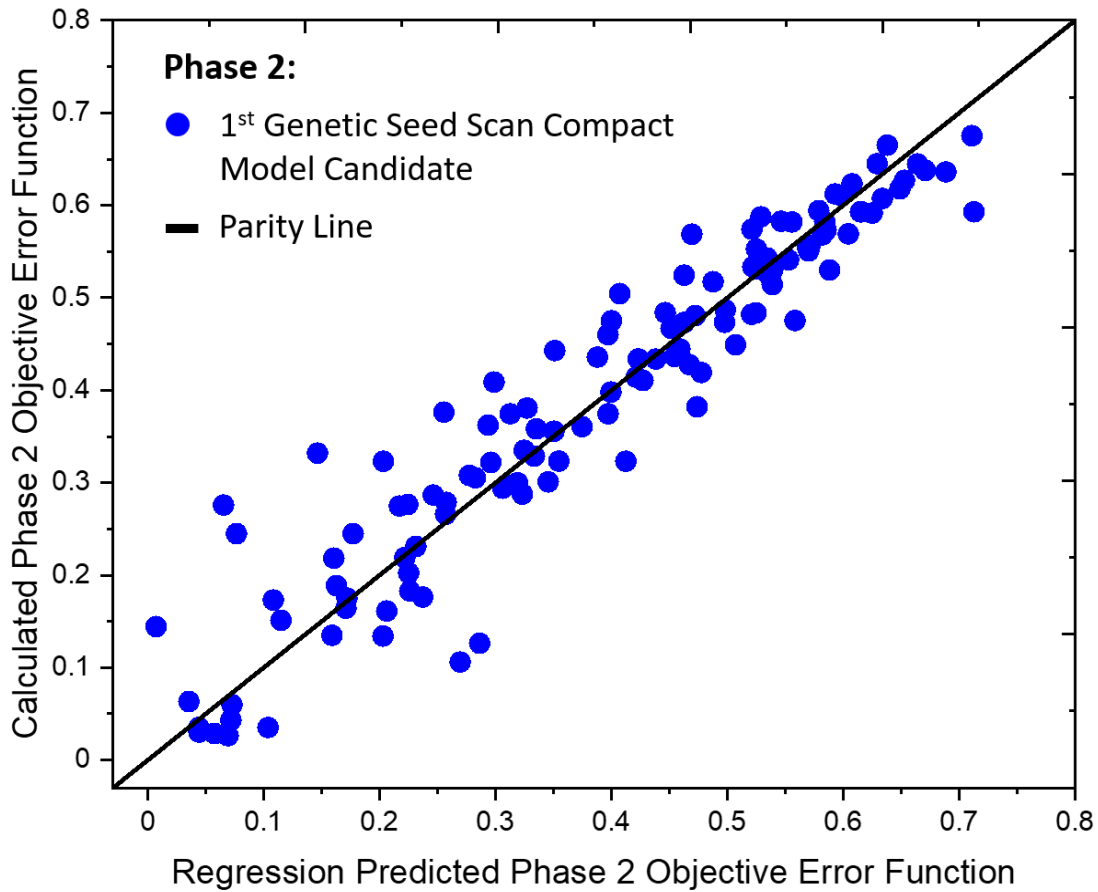

Figure 10. Calculated Phase 2 objective error function (OEF) versus Phase 2 objective error function predicted by multiple linear regression of a subset of compact model candidates produced in the Phase 2 1st Genetic Seed Scan. Phase 2 OEF is comprised of $S_u$ calculations at: 573K, 10 atm, $\phi$ = 0.8, 1.05, 1.3.

**4.3 Effect of the Number of Nodes in the Virtual Reaction Network**

It was previously found that increasing the number of nodes in the virtual reaction network from fifteen to nineteen resulted in a considerably more tractable optimisation task, and a more accurate compact model [21]. The reason for this is that the unoptimised overly-reduced model is not as severely corrupted, meaning the virtual reaction flux must only be altered to a lesser degree than in the case of a fifteen-node virtual reaction network. Therefore, the optimal kinetic parameters tend to possess values closer to the original values, requiring a smaller search space and resulting in a less complex OEF topology. Furthermore, with more nodes in the virtual reaction network there exists more connections (degrees of freedom) for the virtual reaction flux to be allocated to, resulting in a high-fidelity description of the detailed model behaviours, most notably at low temperature and fuel-rich conditions.

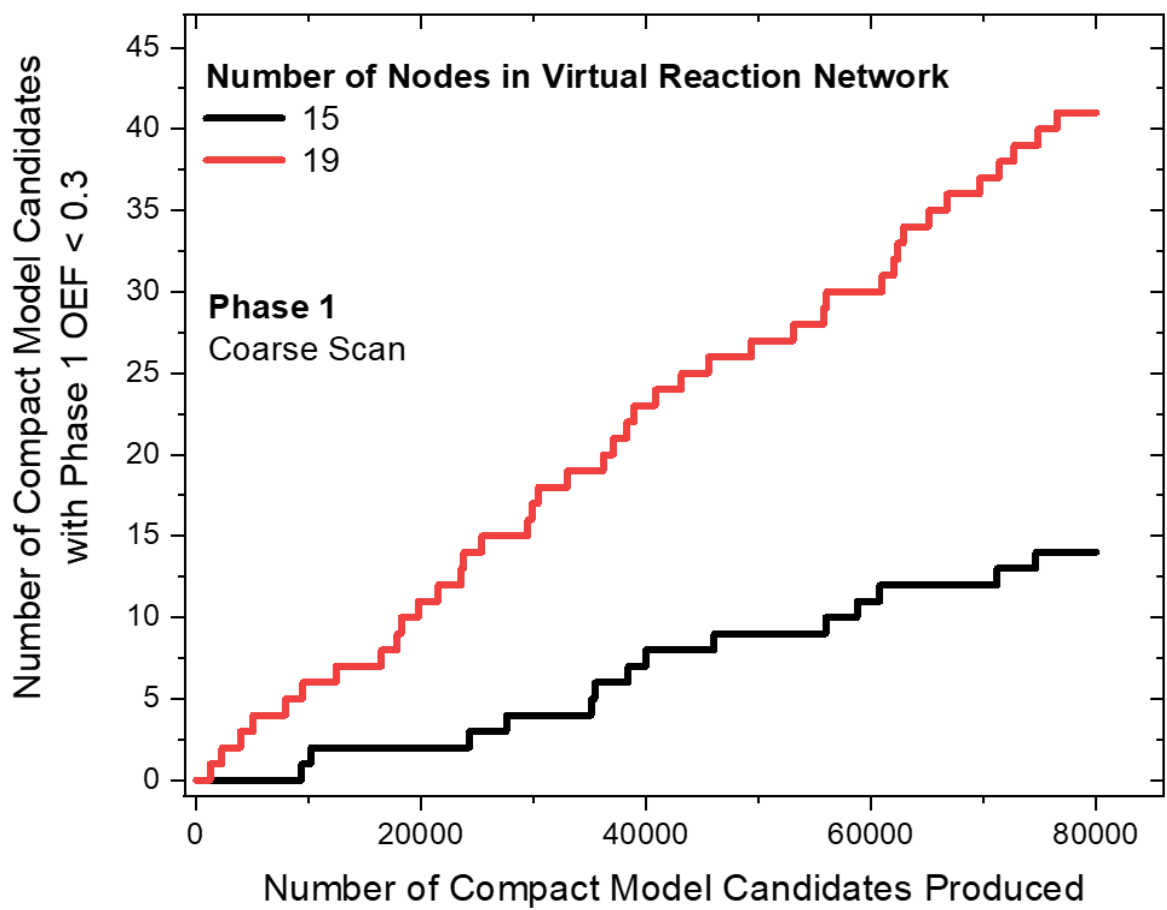

Figure 11. The number of compact model candidates produced with a Phase 1 objective error function (OEF) less than 0.3 in the Coarse Scan during 15- and 19-node compaction, as a function of the number of compact model candidates produced in the scan. The compact model candidate with the lowest Phase 1 OEF was 0.21 for 15 node and 0.12 in the case of 19 node compaction.

While a nineteen-node compact model was not produced in this work, a Phase 1 Coarse Scan was performed on a nineteen-node reaction network construct to highlight the difference in behaviour. Figure 11 demonstrates this by counting the number of compact model candidates produced in the Phase 1 Coarse Scan with a 0-D OEF below 0.3 in the case of fifteen and nineteen node compaction.

## 5. Conclusions

A novel compute intensification methodology to simplify the description of complex chemical reaction systems was designed and demonstrated on the archetypal example of the methane combustion system. The *Machine Learned Optimisation for Chemical Kinetics (MLOCK)* algorithm allows for the production of minimally complex, high fidelity kinetic models, which we call "Compact Models" (i.e. minimised and optimised). The methodology is intentionally designed to require minimal starting information or other prior human knowledge. Instead, the MLOCK methodology emphasises the leverage of modern computing resources to perform large numbers of complex simulations quickly with minimal user input. The concept of MLOCK is to produce a data set, which is interrogable by objective data analysis methods which can automatically search the data set to find the optimum set (or sets) of numerical terms which results in an accurate replication of the target calculations.

MLOCK is designed to work within the significant constraints imposed by the data science principal of interoperability to the standard reaction kinetic simulation infrastructure of the day (ChemKin format). Even with this restriction, it is shown that the concept readily arrives at a large number of discreet instances of numerical terms that through their combination result in accurate calculation of a broad range of complex chemical kinetic phenomena pertinent to typical multi-dimensional reacting flow gas turbine combustor simulations, as defined by the gas turbine industry.

Specifically, for the simulation of methane combustion for performance target matrix comprising NUIGMECH1.0 detailed model calculations of ignition delay time (IDT), perfectly stirred reactor and laminar flame speed ($S_u$) across a broad range of operating conditions, MLOCK simplifies the reaction network from 2,759 species to a 15-node (species) virtual reaction network while retaining fidelities in the range of 87-89% relative to the calculations of the detailed model in the simulation of a broad range of 0-D and 1-D methane combustion kinetic calculations, at the broad range of conditions of relevance to the gas-turbine industry. This results in a compact model with a computational time to solve a 1-D laminar flame of 60% that of DRM22 whilst retaining a similar degree of accuracy.

This work demonstrates the potential of the coupling of data intensive and machine learning methodologies with the concept of virtual descriptions of reaction networks as a means to accurately and cost-effectively describe physical phenomena owing to the occurrence of complex chemical reaction mechanisms.

## Acknowledgements

The research reported in this publication was support by funding from Siemens Canada Limited and the Sustainable Energy and Fuel Efficiency (SEFE) Spoke of MaREI, the SFI Centre for Energy, Climate and Marine Research (16/SP/3829). Mark Kelly is supported by the Irish Research Council (GOIPG/2020/1043). Calculations were performed on the Boyle cluster maintained by the Trinity Centre for High Performance Computing.